\documentclass[
prd
,showpacs,amssymb,superscriptaddress,aps
]{revtex4-2}
\usepackage{graphicx}
\usepackage{color}
\input{epsf}

\usepackage{amsmath,amssymb}
\usepackage{bm}
\usepackage{times}
\usepackage{ulem}

\newcommand{\dalm}{\kern1pt\vbox{\hrule height 0.9pt\hbox{\vrule width 0.9pt
\hskip 2.5pt\vbox{\vskip 5.5pt}\hskip 3pt\vrule width 0.3pt}\hrule height 0.3pt}
\kern1pt}

\begin{document}



\title{Shear oscillations in neutron stars and the nuclear symmetry energy}

\author{Hajime Sotani}
\email{sotani@yukawa.kyoto-u.ac.jp}
\affiliation{Astrophysical Big Bang Laboratory, RIKEN, Saitama 351-0198, Japan}
\affiliation{Interdisciplinary Theoretical \& Mathematical Science Program (iTHEMS), RIKEN, Saitama 351-0198, Japan}



\date{\today}

\begin{abstract}
The shear and interface modes excited inside the neutron star due to the presence of elasticity depend on the properties of both the crust and core region. To examine how such eigenfrequencies depend on the neutron star properties, we solve the eigenvalue problem by adopting the relativistic Cowling approximation. Then, we confirm that the number of the interface modes excited in the star is generally equivalent to the number of the interface where the shear modulus discontinuously becomes zero, but we also find that the number of interface modes becomes smaller than that of the interface for the stellar model with lower or higher value of the slope parameter $L$. Furthermore, we derive the empirical relations for expressing the shear modes and one of the interface modes ($i_1$-mode in the text), which is the mode whose amplitude becomes dominant at the interface between the crust and envelopes and at the interface between the phases of slablike and cylindrical nuclei. At the end, we also show the possibility of identifying the higher QPO frequencies observed in GRB 200415A with the shear oscillations, as an alternative possibility instead of the torsional oscillations. 
\end{abstract}

\pacs{04.40.Dg, 97.10.Sj, 04.30.-w, 26.60.Gj}
%
\maketitle


\section{Introduction}
\label{sec:I}

Neutron stars are one of the most suitable environments for probing physics under extreme states. In fact, the density inside the neutron star easily exceeds the nuclear saturation density \cite{ST83}. Through the observation of neutron stars, one might touch the aspect in such a high-density region, which is quite difficult to realize on Earth due to the nature of nuclear saturation properties. For example, the discoveries of massive neutron stars, whose masses are around $2M_\odot$ \cite{D10,A13,C20,F21, R22}, have excluded soft equations of state (EOSs), with which the expected maximum mass of neutron stars is less than the observations. The gravitational waves from binary neutron star merger, GW170817 \cite{gw170817}, tell us the tidal deformability of neutron stars, which leads to the constraint on the $1.4M_\odot$ neutron star radius \cite{Annala18}. Moreover, since the light radiating from a neutron star's surface bends due to a strong gravitational field induced by the neutron star as a relativistic effect, one could predominantly constrain the stellar compactness, which is the ratio of the mass to the radius, through careful observations of the pulsar light curve (e.g., \cite{PFC83,LL95,PG03,PO14,SM18,Sotani20a}). In practice, the Neutron star Interior Composition Explorer (NICER) operating on an International Space Station could constrain the neutron star mass and radius for PSR J0030+0451 \cite{Riley19,Miller19} and for PSR J0740+6620 \cite{Riley21,Miller21}. In addition to the astronomical observations, using the terrestrial experiments, one may constrain the EOS for neutron star matter in a relatively lower density region (e.g., Refs. \cite{SNN22,SO22,SN23}).

The frequencies from a neutron star, if observed, are another important piece of information to extract the neutron star properties. Since the specific oscillation modes excited in the objects strongly depend on their interior properties, one could extract the properties through the observation of the corresponding frequencies as an inverse problem. This technique is known as asteroseismology, which is similar to seismology on Earth or helioseismology on Sun. In practice, by identifying the quasi-periodic oscillations (QPOs) observed in the magnetar giant flares with the crustal torsional oscillations, one could constrain the stellar properties (e.g., Refs. \cite{SW2009,GNHL2011,SNIO2012,SIO2016,SIO2019,SKS23}). In a similar way, once the gravitational waves from an (old) neutron star would be observed, one may extract the stellar radius, mass, and EOS (e.g., \cite{AK1996,AK1998,STM2001,SH2003,SYMT2011,PA2012,DGKK2013,KHA2015,Sotani2020,Sotani21,SD2021}).

Several eigenmodes can be excited simultaneously on neutron stars due to various physical processes, among which we particularly focus on the shear ($s$-) and interface ($i$-) modes in this study. Both modes are excited due to the presence of elasticity inside the star \cite{KHA2015,Finn90,SL02,PB05,Sotani23}. In this study, we assume that the core region (and envelope) of the neutron star behaves as a fluid, while the elasticity is present only inside the crust region. From an observational point of view, elastic oscillations may become important to explain the precursors and QPOs in precursors observed just before the main flare activity of gamma-ray burst from a binary neutron star merger \cite{Troja10,Xiao22}, which are considered as a result of resonant shattering of neutron star crusts induced by the binary orbital motion \cite{Tsang12,Tsang13}. In previous our study, we have shown that the $i$-mode frequencies, $f_i$, multiplied with stellar mass, $M$, and $s$-mode frequencies, $f_s$, multiplied with stellar radius, $R$, are well expressed as a function of stellar compactness independently of the stiffness of the core region, which may enable us to extract the crustal information, once one would observe the $i$- or $s$-mode frequencies. However, we adopted only two EOSs in the previous study and could not discuss the dependence of the frequencies on the nuclear saturation parameters. Thus, in this study, we adopt several EOSs with different saturation parameters and will discuss how the frequencies can be characterized by the nuclear saturation parameters.

This manuscript is organized as follows. In Sec. \ref{sec:EOS}, we briefly mention the equilibrium neutron star models, the EOS considered in this study, and the shear modulus inside the neutron star crust. In Sec. \ref{sec:Eigen}, we focus on the $i$-modes excited in the neutron stars. In Sec.~\ref{sec:ER}, we derive the empirical relations for the frequencies of the $i$- and $s$-mode oscillations. Furthermore, in Sec. \ref{sec:identification} we also discuss the possibility for identification of higher frequency QPOs observed in GRB 200415A \cite{CT21} with the shear oscillations. Finally, we conclude this study in Sec. \ref{sec:Conclusion}. Unless otherwise mentioned, we adopt geometric units in the following, $c=G=1$, where $c$ and $G$ denote the speed of light and the gravitational constant, respectively.

\section{EOS and Equilibrium models}
\label{sec:EOS}

In a similar way in Ref.~\cite{Sotani23}, we simply consider a non-rotating, strain-free, and spherically symmetric neutron star as an equilibrium model in this study. The metric describing such an object is given by
\begin{equation}
  ds^2 = -e^{2\Phi}dt^2 + e^{2\Lambda}dr^2 + r^2\left(d\theta^2 + \sin^2\theta d\phi^2\right), \label{eq:metric}
\end{equation}
where $\Phi$ and $\Lambda$ are the metric functions depending on only $r$. The mass function, $m(r)$, which is the enclosed (gravitational) mass inside the position $r$, is associated with $\Lambda$ through $e^{-2\Lambda}=1-2m/r$. Once an appropriate EOS is selected, one can construct the stellar models by integrating the Tolman-Oppenheimer-Volkoff equation. Since we focus on the shear oscillations excited by the crust elasticity, the position of the crust surface is crucial. In general, one can neglect the thermal effect on the neutron structure, because the Fermi temperature inside the neutron star becomes much higher than the physical temperature. However, this condition may break in the vicinity of the stellar surface. In fact, as the density decreases, the Fermi temperature also decreases and eventually becomes comparable to or less than the physical temperature. As a result, the density of the crust surface strongly depends on the radial distribution of physical temperature \cite{GPE83}. Nevertheless, as in Ref.~\cite{Sotani23}, we simply assume that the density of the crust surface is $10^{10}$ g/cm$^3$ and the surface density is $10^6$ g/cm$^3$, which are typical values (see Appendix \ref{sec:appendix_1} for the dependence of the $i$-mode frequencies on the surface density).

Regarding the EOS, we particularly adopt the phenomenological EOS in a non-relativistic framework proposed by Oyamatsu and Iida~\cite{OI03,OI07} (hereafter referred to as OI-EOSs). The energy per nucleon, $E/A$, for any EOSs can be expanded around the saturation density $n_0$ for symmetric nuclear matter as a function of the baryon number density, $n_{\rm b}$, and an asymmetry parameter, $\alpha$, as
\begin{equation}
  \frac{E}{A} = w_0 + \frac{K_0}{2}u^2 + {\cal O}(u^3) 
     + \alpha^2 \left[S_0 + Lu + {\cal O}(u^2)\right] + {\cal O}(\alpha^3), \label{eq:E/A}
\end{equation}
where $n_{\rm b}=n_n+n_p$ and $\alpha=(n_n-n_p)/n_{\rm b}$ with the neutron number density, $n_n$, and the proton number density, $n_p$, while $u=(n_{\rm b}-n_0)/(3n_0)$. In this expression, the five coefficients ($n_0$, $w_0$, $K_0$, $S_0$, and $L$) are the (least) nuclear saturation parameters, depending on the adopted EOS. That is, each EOS has its own set of nuclear saturation parameters, which characterizes the EOS. 
The parts without the $\alpha$ dependence correspond to the energy per nucleon for a symmetric nuclear matter, while the coefficient of $\alpha^2$ corresponds to the symmetry energy, $S(u)$. The saturation density, $n_0$, for a symmetric nuclear matter is determined by $\partial (E/A) /\partial u=0$, assuming $\alpha=0$. $w_0$ denotes the binding energy for the symmetric nuclear matter at $n_{\rm b}=n_0$, while $K_0$, the so-called incompressibility, is the second derivative concerning $u$ for symmetric nuclear matter. Meanwhile, $S_0$ is the symmetry energy at $n_{\rm b}=n_0$ and $L$ is the density dependence of $S(u)$, i.e., $L\equiv \partial S(u)/\partial u$.
We remark that the Skyrme-type EOSs and the EOSs based on the relativistic mean field (RMF) theory are originally constructed with the interaction involving only up to the terms of $\alpha^2$. The EOSs beyond the Skyrme-type EOSs or the RMF EOSs may contain more than the $\alpha^4$ (or $\alpha^3$) terms, but the contribution of the $\alpha^4$ terms may be tiny, e.g., the deviation from the expression up to the $\alpha^2$ terms given by Eq.~(\ref{eq:E/A}) is only $\sim 1\%$ \cite{TT13}. 

Among these five parameters, $n_0$, $w_0$, and $S_0$ are relatively well constrained from the nuclear experiments, while the constraint on $K_0$ and $L$ is more difficult because one needs to know the experimental data in a wide density range to determine the density derivative. We note that the current constraint on $K_0$ and $L$ are $K_0=240\pm 20$ MeV \cite{Sholomo} and $L=60\pm 20$ MeV \cite{Vinas14,Li19}. 
As one can expect from the expression of $E/A$, $K_0$ and $L$ strongly affect the stiffness of a  neutron star EOS, especially for a higher-density region. That is, an EOS with larger $K_0$ and/or $L$ becomes a stiffer EOS in a higher-density region. On the other hand, $K_0$ and $L$ also affect the crust properties, which correspond to the density region lower than the saturation density. Since the symmetry energy in a density region lower than the saturation density becomes small with an EOS with larger  $K_0$ and $L$, it is easier for the proton to change into a neutron. That is, the ion charge number would be smaller for the EOS with larger $K_0$ and $L$. This tendency is important to discuss the elastic oscillations because the shear modulus in the phase of spherically symmetric nuclei is proportional to the square of the ion charge (see Eq.~(\ref{eq:musp})).
Anyway, in order to systematically examine the saturation parameter dependence of the neutron star oscillation frequencies, we adopt the OI-EOS family in this study. For the given values of $K_0$ and $L$, OI-EOSs are constructed in such a way that the values of $n_0$, $w_0$, and $S_0$ are tuned by recovering the empirical nuclear data with the extended Thomas-Fermi theory. The EOS parameters adopted in this study are listed in Table~\ref{tab:EOS}, where we also list the transition density from spherical to cylindrical nuclei (SP-C), from cylindrical to slablike nuclei (C-S), from slablike to cylindrical-hole nuclei (S-CH), from cylindrical-hole to spherical-hole nuclei (CH-SH), and from spherical-hole nuclei to uniform matter (SH-U) for each EOS. We note that OI-EOSs with $(K_0,L)=(230,42.6)$ and $(230,73.4)$ are the same EOS models adopted in Ref.~\cite{Sotani23}. We also note that some of the EOS parameters shown in Table~\ref{tab:EOS} are out of the current constraints on $K_0$ and $L$, but we consider the EOSs listed in Table~\ref{tab:EOS} to examine the EOS parameter dependence in a wide parameter region.

\begin{table*}
\centering
\caption{
The EOS parameters adopted in this study. SP-C, C-S, S-CH, CH-SH, and SH-U denote the transition densities in the OI-EOSs characterized by $K_0$ and $L$. 
}
\begin{tabular}{cccccccc}
\hline\hline
  $K_0$ (MeV)  & $L$ (MeV) & SP-C (fm$^{-3}$) & C-S (fm$^{-3}$) & S-CH (fm$^{-3}$) & CH-SH (fm$^{-3}$) & SH-U (fm$^{-3}$)  \\
\hline
  180 & 31.0 & 0.05887 & 0.07629 & 0.08739 & 0.09000   & 0.09068       \\  
  180 & 52.2 & 0.06000 & 0.07186 & 0.07733 & 0.07885   & 0.07899    \\  
  230  & 23.7 & 0.05957 & 0.07997 & 0.09515 & 0.09817  &  0.09866        \\  
  230 & 42.6 & 0.06238 & 0.07671 & 0.08411 & 0.08604   & 0.08637  \\   
  230 & 73.4 & 0.06421 & 0.07099 & 0.07284 & 0.07344   & 0.07345  \\  
  360  & 40.9 & 0.06743 & 0.08318 & 0.09197 & 0.09379  &  0.09414       \\  
\hline\hline
\end{tabular}
\label{tab:EOS}
\end{table*}

Furthermore, to see the dependence of the EOS stiffness in a higher-density region, we consider not only the original OI-EOSs but also the original OI-EOSs for  $\varepsilon\le\varepsilon_t$ connected to a one-parameter EOS for  $\varepsilon\ge\varepsilon_t$ expressed as
\begin{equation}
  p = \alpha(\varepsilon-\varepsilon_t) + p_t, \label{eq:eos_h}
\end{equation}
where $p_t$ is the pressure at $\varepsilon=\varepsilon_t$ given from the OI-EOSs. We note that $\alpha$ corresponds to the square of the sound velocity. Since the EOSs are expressed more or less using their own nuclear saturation parameters up to twice the saturation density (e.g.,  \cite{SIOO14}), we assume that $\varepsilon_t$ is equivalent to twice the saturation density in this study, as in Refs.~\cite{Sotani17,Sotani23}. From Table~\ref{tab:EOS}, one can observe that the density at the boundary between the crust and core, i.e., SH-U, is at most $\sim 2n_0/3$, considering that $n_0\approx 0.15-0.16$ fm$^{-3}$ \cite{OHKT17}. The crust thickness depends on $L$ (and $K_0$) and stellar compactness, $M/R$, where the thickness decreases as $L$ and $M/R$ increases \cite{OI07,SIO2017}. In this study, we especially focus on the case of $\alpha=1/3, 0.6,$ and 1, as in Ref. \cite{Sotani23}.
In this study, we simply discuss the stiffness of the core region of a neutron star by assuming the simple EOS given by Eq. (\ref{eq:eos_h}) with the OI-EOSs constructed in a non-relativistic formalism. However, since the properties in a higher-density region generally depend on whether the EOS is based on a non-relativistic or relativistic formalism, it may be important to see the dependence on the unified EOSs constructed with different formalisms. The unified EOSs including the pasta phases are unfortunately quite few now, but we will examine such dependence in the future.

The shear modulus, $\mu$, is another important property characterizing the elasticity.
As mentioned above, neutron stars have a lot of eigenfrequencies, depending on input physics. By introducing the crust elasticity, the additional eigenmodes, i.e., the $i$- and $s$-modes in the polar-parity oscillations or the torsional ($t$-) modes in the axial-parity oscillations, are excited. This is a reason why the $i$- and $s$-modes considered in this study are important to extract the crust properties and also why we have to explicitly define the shear moduli here.
The shear modulus, $\mu_{\rm sp}$, in the phase composed of spherical nuclei has been studied relatively well, but here we adopt the standard shear modulus formulated as a function of ion charge number, $Z$, ion number density, $n_i$, and Wigner-Seitz cell radius, $a$, which is related to $n_i$ through $4\pi a^3/3=1/n_i$:
\begin{equation}
  \mu_{\rm sp} = 0.1194\frac{n_i(Ze)^2}{a}, \label{eq:musp}
\end{equation}
assuming that the spherical nuclei form the body-centered cubic (bcc) lattice with a point-like ion \cite{SHOII1991}. We note that the shear modulus could be modified a little with the phonon contribution \cite{Baiko2011}, the electron screening effect \cite{KP2013}, the polycrystalline effect \cite{KP2015}, and the effect of finite sizes of atomic nuclei \cite{STT22b}. The shear modulus, $\mu_{\rm cy}$, in the phase composed of cylindrical nuclei and the shear modus, $\mu_{sl}$, in the phase composed of slablike nuclei have been also discussed in Ref.~\cite{PP1998}. According to their study, $\mu_{\rm cy}$ is given by 
\begin{equation}
  \mu_{\rm cy} = \frac{2}{3}E_{\rm Coul} \times 10^{2.1(w_2-0.3)}, \label{eq:mucy}
\end{equation}
as a function of the Coulomb energy per volume of a Wigner-Seitz cell, $E_{\rm Coul}$, and the volume fraction of cylindrical nuclei, $w_2$, while $\mu_{\rm sl}$ is given by 
\begin{equation}
  \mu_{\rm sl} = 0, \label{eq:musl}
\end{equation}
in the linear perturbation level, i.e., the shear elasticity in the phase of slablike nuclei comes from the higher-order perturbations. In this study, we adopt Eqs. (\ref{eq:mucy}) and (\ref{eq:musl}) for estimating $\mu_{\rm cy}$ and $\mu_{\rm sl}$ as in Ref.~\cite{Sotani23}, but recently it has been pointed out that the elastic constant in the polycrystalline lasagna (slablike nuclei) may become a (nonzero) tiny value \cite{CSH18,PZP20}. Moreover, the shear modulus, $\mu_{\rm ch}$ $(\mu_{\rm sh})$, in the phase of cylindrical-hole (spherical-hole) nuclei can be estimated, using the formulas for $\mu_{\rm cy}$ ($\mu_{\rm sp}$), because the structure of liquid crystalline composed of cylindrical-hole (spherical-hole) nuclei are the same as that of cylindrical (spherical) nuclei (see Ref.~\cite{SIO2019} for details).


\section{Eigenfunctions of the $i$-modes}
\label{sec:Eigen}

To determine the specific frequencies of neutron stars, we make a linear perturbation analysis on the equilibrium models mentioned in the previous section. For this purpose, we adopt the relativistic Cowling approximation in this study, where the metric perturbations are neglected during the fluid oscillation. The perturbation equations are derived from linearizing the energy-momentum conservation law. By imposing the appropriate boundary conditions and junction conditions, the problem to solve becomes the eigenvalue problem. 
Since the perturbation equations, boundary conditions imposed at the stellar center and the surface, and the junction conditions imposed at the interface where the shear modulus discontinuously becomes zero are completely the same as shown in Refs.~\cite{SL02,Sotani23}, we avoid explicitly showing them here. But, we briefly mention how to determine the eigenvalues, using such perturbation equations, boundary conditions, and junction conditions. The perturbation equations are integrated outward from the stellar center with the boundary conditions, assuming a trial value of eigenvalue, $\omega$, up to the interface between the core and the phase of spherical-hole nuclei. The resultant perturbative variables at the interface are changed into those for the elastic region with the junction condition, then they are integrated outward again up to the next interface, which corresponds to the position between the phases of cylindrical-hole and slab-like nuclei. By iterating these procedures, finally, the perturbative variables at the stellar surface are determined. Since the resultant perturbative variables do not generally satisfy the boundary conditions at the surface, we have to search the suitable value of $\omega$, with which the boundary conditions at the surface satisfy, by changing $\omega$. After we find the value of $\omega$ in such a way, the eigenfrequency, $f$, is determined via $f=\omega/(2\pi)$.
In this study, we focus only on the $\ell=2$ oscillation modes.

The $i$-modes are excited due to the existence of the interface between the phases with zero and nonzero shear modulus \cite{PB05}. Thus, in the most general case, the number of the $i$-modes excited in a neutron star becomes equivalent to the number of the interfaces, where the non-zero elasticity discontinuously becomes zero. That is, since our model has four interfaces inside a neutron star composed of a liquid core, a solid crust with pasta structures, and envelopes, i.e., the boundary between the envelopes and the surface of curst (the phase of spherical nuclei), the boundary between the phases of cylindrical and slablike nuclei, the boundary between the phases of slablike and cylindrical-hole nuclei, and the boundary between the phase of spherical-hole nuclei and core, one can expect the excitation of four $i$-modes. Nevertheless, we have found that only three $i$-modes can be excited in the neutron star model with a relatively larger value of $L$. This may be because the thickness of the phases composed of cylindrical-hole and spherical-hole nuclei is relatively narrow, which makes a situation that one of four $i$-modes is difficult to be excited \cite{Sotani23}. So, in this section, we carefully see the behavior (and especially the number) of the $i$-mode frequencies.

\begin{figure*}[tbp]
\begin{center}
\includegraphics[scale=0.4]{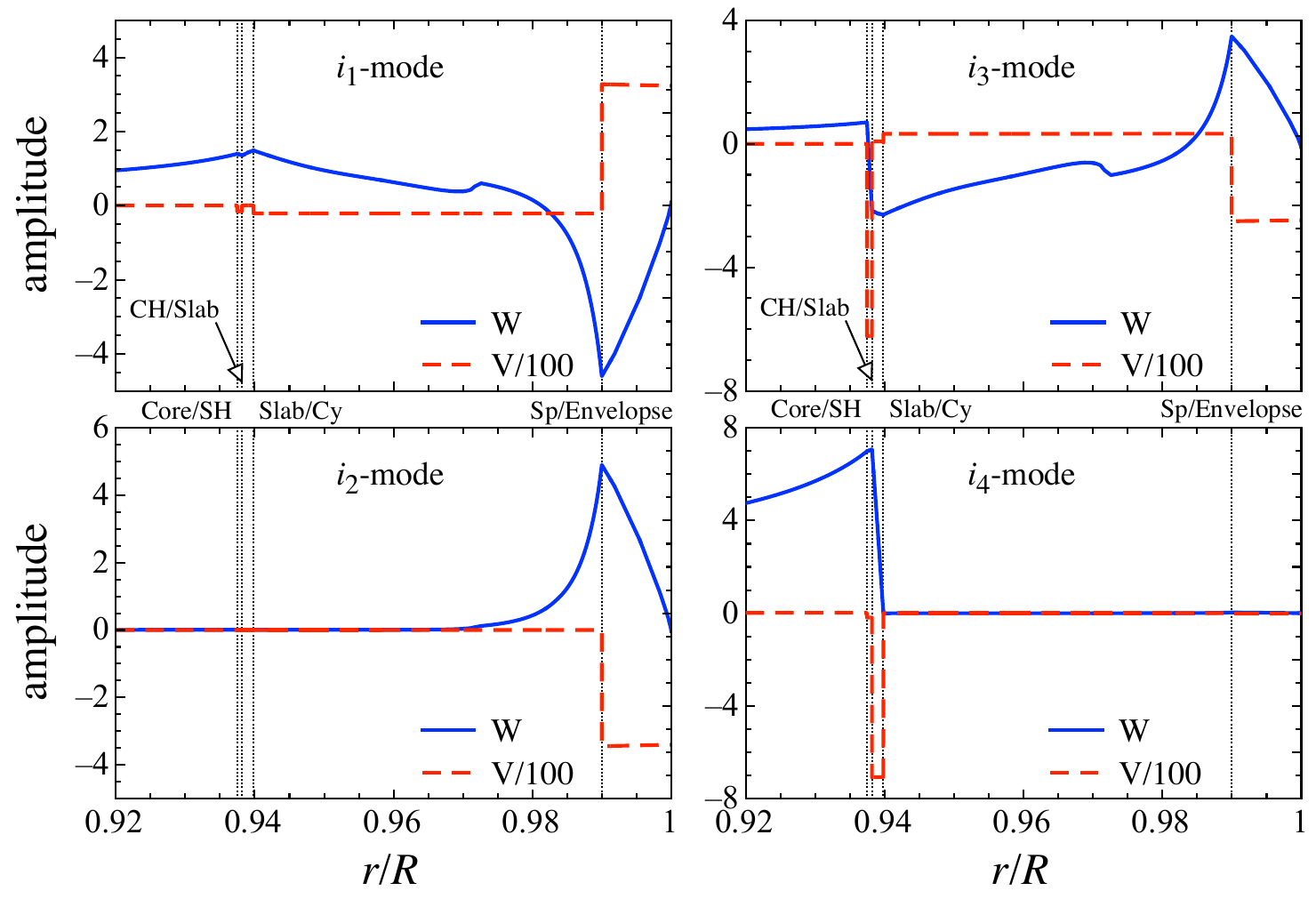}
\end{center}
\caption{
Radial profile of the amplitude of $i_i$-modes for the neutron star mode with $1.44M_\odot$ and 10.2 km using the EOS with $K_0=230$ and $L=42.6$ MeV.
Here, $W$ and $V$ denote the Lagrangian displacement in the radial and angular directions.
The vertical lines from left to right denote the boundaries between core and spherical-hole nuclei, between cylindrical-hole and slablike nuclei, between slablike and cylindrical nuclei, and between spherical nuclei and envelop. The eigenfunctions shown in the left panels correspond to the modes excited even without the pasta structure, while those in the right panels correspond to the modes additionally excited due to the presence of the phase composed of cylindrical-hole and spherical-hole nuclei.
}
\label{fig:imode}
\end{figure*}

In Fig.~\ref{fig:imode}, as an example of a typical case, we show the radial profile of the eigenfunction of $i_i$-modes for the neuron star model with $1.44M_\odot$ and 10.2 km, using the EOS with $K_0=230$ and $L=42.6$ MeV, where $W$ and $V$ denote the Lagrangian displacement in the radial and angular directions. In each panel, we show four vertical dotted lines, which correspond to the position of the interface, i.e., the boundaries between core and spherical-hole nuclei, between cylindrical-hole and slablike nuclei, between slablike and cylindrical nuclei, and between spherical nuclei and envelop from left to right. The top-left and bottom-left panels correspond to the $i_1$- and $i_2$-modes, which are dominantly excited inside the phase composed of spherical and cylindrical nuclei, while the top-right and bottom-right panels correspond to the $i_3$- and $i_4$-modes. One can observe that the $i_3$-modes are excited even inside the phases of spherical-hole and cylindrical-hole nuclei, while the $i_4$-modes are inside the phase of slablike nuclei. We note that the functional form for the $i_1$-, $i_2$-, and $i_3$-modes is essentially the same as shown in Figs.~4 and 5 in Ref.~\cite{Sotani23}. In this study, we identify the $i$-mode by checking these features of each eigenfunction.

\begin{figure*}[tbp]
\begin{center}
\includegraphics[scale=0.45]{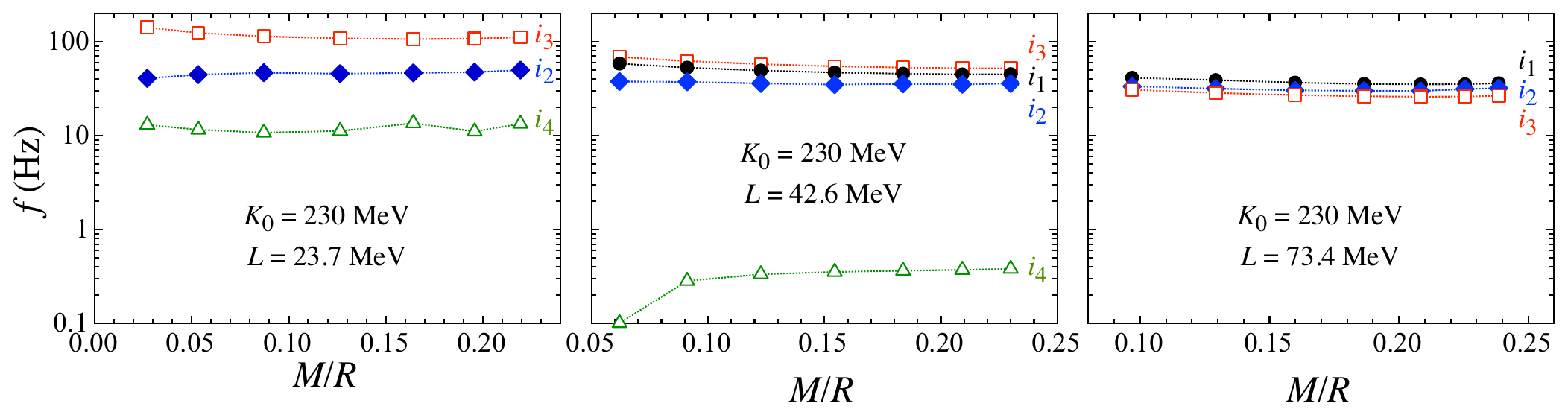}
\end{center}
\caption{
Eigenfrequencies of the $i_i$-modes are shown as a function of the stellar compactness, $M/R$, for the neutron star models constructed using the original OI-EOSs with $K_0=230$ MeV and $L=23.7, 42.6$, and $73.4$ MeV in the left, middle, and right panel, respectively. We note the labeling of the $i_i$-modes for the stellar model shown in the middle panel is different from that in the previous study \cite{Sotani23} (see text for details).
}
\label{fig:yMR}
\end{figure*}

\begin{table*}
\centering
\caption{The order of $i_i$-modes excited in the neutron stars, depending on the EOS parameters. In general, the $i_i$-modes excited in the neutron star are in order of $i_4$, $i_2$, $i_1$, and $i_3$ from low to high frequencies, but the order becomes different with some EOS models with higher or lower $L$, which are especially shown with bold font. 
}
\begin{tabular}{cc|cccc}
\hline\hline
  $K_0$ (MeV)  & $L$ (MeV) & original OI  & $\alpha=1/3$ & $\alpha=0.6$ & $\alpha=1$   \\
\hline
  180 & 31.0 & $\bm {(i_4, i_2, i_3)}$ & ($i_4$, $i_2$, $i_1$, $i_3$) & ($i_4$, $i_2$, $i_1$, $i_3$) & ($i_4$, $i_2$, $i_1$, $i_3$)   \\  
  180 & 52.2 & ($i_4$, $i_2$, $i_1$, $i_3$) & ($i_4$, $i_2$, $i_1$, $i_3$) & ($i_4$, $i_2$, $i_1$, $i_3$) & ($i_4$, $i_2$, $i_1$, $i_3$)     \\  
  230  & 23.7 & $\bm {(i_4, i_2, i_3)}$ & ($i_4$, $i_2$, $i_1$, $i_3$) & ($i_4$, $i_2$, $i_1$, $i_3$) & ($i_4$, $i_2$, $i_1$, $i_3$)     \\  
  230 & 42.6 & ($i_4$, $i_2$, $i_1$, $i_3$) & ($i_4$, $i_2$, $i_1$, $i_3$) & ($i_4$, $i_2$, $i_1$, $i_3$) & ($i_4$, $i_2$, $i_1$, $i_3$)      \\   
  230 & 73.4 & $\bm {(i_3, i_2, i_1)}$ & $\bm {(i_3, i_2, i_1)}$ & $\bm {(i_3, i_2, i_1)}$ &  $\bm {(i_3, i_2, i_1)}$     \\  
  360  & 40.9 & ($i_4$, $i_2$, $i_1$, $i_3$) & ($i_4$, $i_2$, $i_1$, $i_3$) & ($i_4$, $i_2$, $i_1$, $i_3$) & ($i_4$, $i_2$, $i_1$, $i_3$)      \\  
\hline\hline
\end{tabular}
\label{tab:modes}
\end{table*}

In the previous study, we simply assigned the $i_i$-modes in order from the highest to the lowest frequencies. But, considering the shape of the eigenfunctions of $i_i$-modes shown in Fig.~\ref{fig:imode}, it is found that the $i_i$-modes should be assigned as the $i_3$, $i_1$, $i_2$, and $i_4$-modes from the highest to the lowest frequencies,  when four $i$-modes are excited (as shown in the middle panel of Fig.~\ref{fig:yMR}). In Fig.~\ref{fig:yMR}, the $i_i$-modes are shown as a function of the stellar compactness for the neutron star models constructed using the original OI-EOS with $K_0=230$ MeV and $L=23.7$, $42.6$, and $73.4$ MeV in the left, middle, and right panel, respectively. From this figure, we find that the stellar models constructed with not only a larger value of $L$ but also a smaller value of $L$ may have only three $i$-modes. We also find that the $i_1$-mode seems to disappear with a lower value of $L$, while the $i_4$-mode disappears with a higher value of $L$ as shown in the previous study. In fact, in Table~\ref{tab:modes}, we list which modes are excited in the stellar models constructed with various values of nuclear saturation parameters and with different stiffness of core region characterized by $\alpha$. From this result, one can observe that four $i$-modes can be excited generally, as expected. Meanwhile, we also mention that the value of $L$, with which the stellar models have three $i$-modes except for the $i_1$-mode, may be too small, considering the fiducial value of $L=60\pm 20$ MeV \cite{Vinas14,Li19}.

\section{Empirical relations}
\label{sec:ER}

As shown in Ref.~\cite{Sotani23}, the frequencies of the $i_i$- and $s_i$-mode ($f_{i_i}$ and $f_{s_i}$) excited in the neutron star are well expressed independently of the value of $\alpha$ as a function of the stellar compactness as
\begin{gather}
  f_{i_i} M\ ({\rm kHz}/M_\odot)= a_{0i} + a_{1i}x + a_{2i}x^2, \label{eq:fit_i} \\
  f_{s_i} R\ ({\rm kHz\ km}) = b_{0i} + b_{1i}x, \label{eq:fit_s}
\end{gather}
where $x$ denotes the stellar compactness, $M_{1.4}/R_{12}$ with $M_{1.4}\equiv M/1.4M_\odot$ and $R_{12}\equiv R/12 {\rm km}$, and $a_{0i}$, $a_{1i}$, $a_{2i}$, $b_{0i}$, and $b_{1i}$ are the adjusted coefficients depending on the crust properties (or the nuclear saturation parameters). In the following, we will see how the frequencies (strictly speaking $a_{0i}$, $a_{1i}$, $a_{2i}$, $b_{0i}$, and $b_{1i}$) depend on the nuclear saturation parameters.

\subsection{Empirical relations for the $i$-modes}
\label{sec:ERi}

\begin{figure}[tbp]
\begin{center}
\includegraphics[scale=0.5]{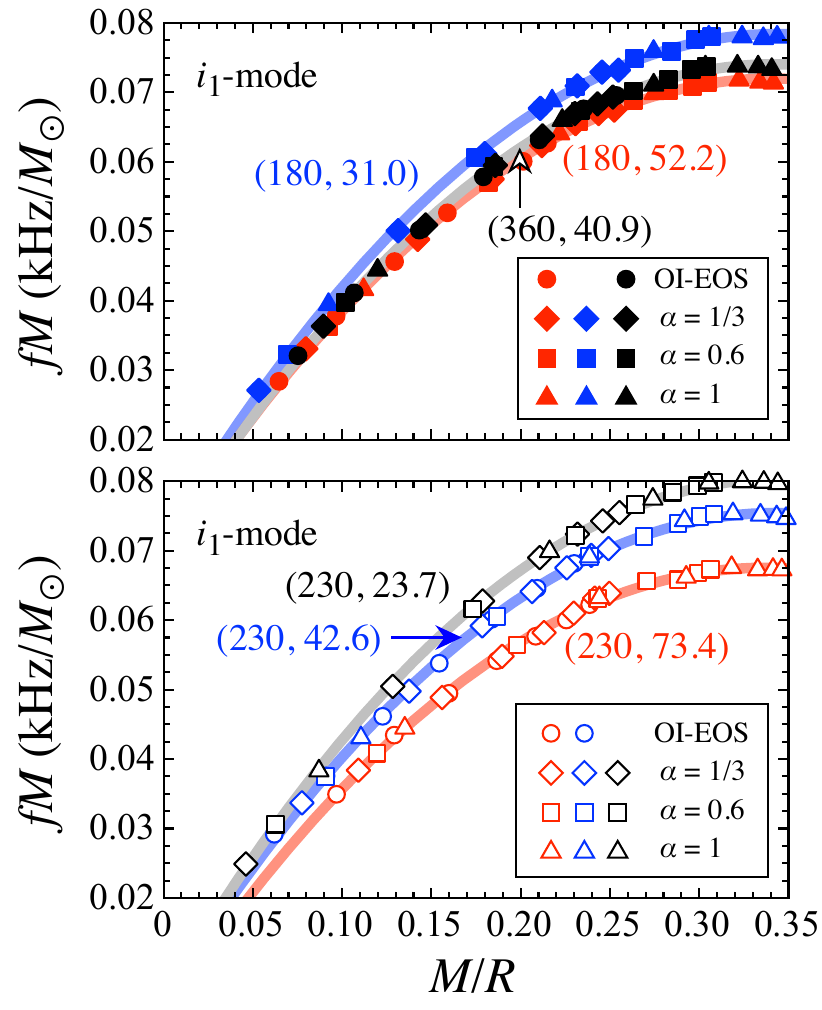}
\end{center}
\caption{
$f_{i_1}M$ for various stellar models are shown as a function of stellar compactness, where the circles, diamonds, squares, and triangles denote the results with stellar models constructed with original OI-EOS, OI-EOS connected to the EOS with $\alpha=1/3$, $0.6$, and $1$, respectively. The solid lines are the fitting given by Eq.~(\ref{eq:fit_i}) and we denote the values of $(K_0, L)$ on each line.
}
\label{fig:i1MR}
\end{figure}

\begin{figure}[tbp]
\begin{center}
\includegraphics[scale=0.5]{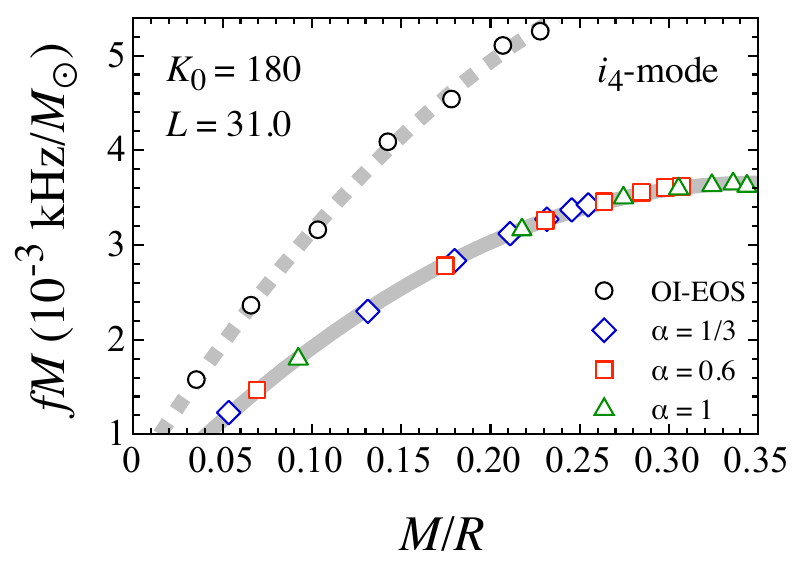}
\end{center}
\caption{
Same as Fig.~\ref{fig:i1MR}, but for $i_4$-mode frequencies on the neutron star models with $K_0=180$ and  $L=31.0$ MeV.
}
\label{fig:i4-y350K180}
\end{figure}

First, we confirm that $f_iM$ is well expressed as a function of $M/R$ given by Eq. (\ref{eq:fit_i}) independently of $\alpha$ (the stiffness of stellar core) as shown in Fig.~\ref{fig:i1MR}, where the solid lines denote the fitting by Eq. (\ref{eq:fit_i}), while various marks correspond to the values of $f_{i1}M$ for the stellar models constructed with different stiffness of core region. On the other hand, we also find an exceptional instance in the $i_4$-modes for the stellar models having three $i$-modes except for the $i_1$-mode with $(K_0, L)=(180, 31.0)$ and $(230, 23.7)$, although we can still confirm that the $i_4$-modes are expressed with Eq. (\ref{eq:fit_i}) independently of $\alpha$ as in Fig.~\ref{fig:i1MR} for the stellar models having the four $ i$-modes. That is, as shown in Fig. \ref{fig:i4-y350K180} for the stellar model with $(K_0, L)=(180, 31.0)$, the $i_4$-mode in the stellar model with the original OI-EOS is obviously different behavior from those with $\alpha=1/3$, 0.6, and 1. In fact, as shown in the right panel of Fig.~\ref{fig:yMR}, the $i_4$-mode frequency seems to appear much higher for the stellar model without the $i_1$-mode. Anyway, considering the fiducial value of $L$ constrained from the experiments, such as $L=60\pm20$ MeV, an exceptional instance we found may not be realized.

\begin{figure*}[tbp]
\begin{center}
\includegraphics[scale=0.33]{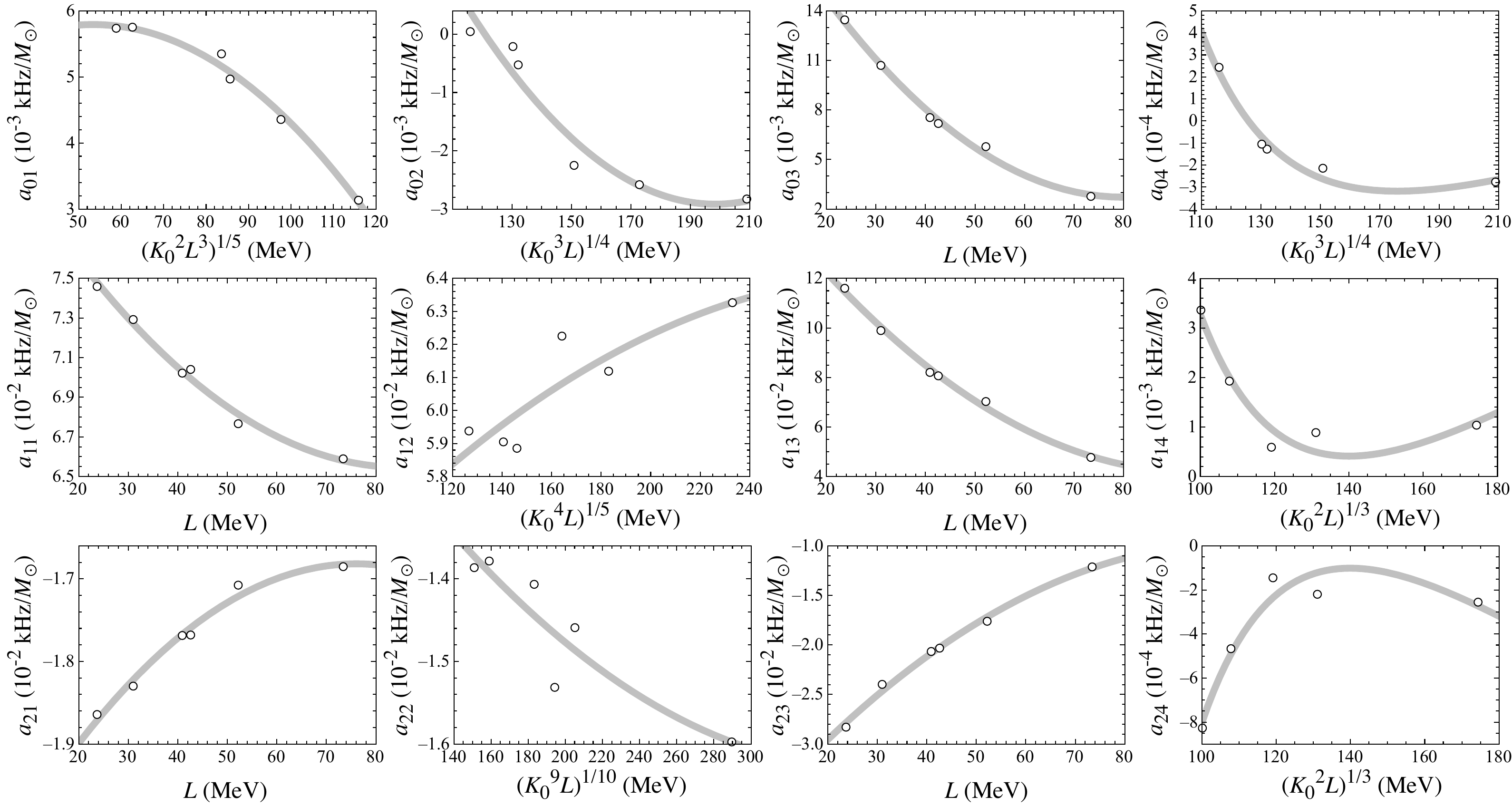}
\end{center}
\caption{
Correlation between the coefficients in Eq.~(\ref{eq:fit_i}) and $L$ or the combination of $K_0$ and $L$, where the solid lines denote the fitting given by Eqs. (\ref{eq:aji}) and (\ref{eq:aj4}). The panels from left to right correspond to $a_{ji}$ for $i=1,2,3,4$, respectively. 
}
\label{fig:aji}
\end{figure*}

Then, we will see the dependence of the $i$-mode frequencies on the nuclear saturation parameters. 
The coefficients in Eq.~(\ref{eq:fit_i}), i.e., $a_{ji}$ for $j=0,1,2$ and $i=1\sim 4$, depend on the crustal properties characterized by the nuclear saturation parameters. Through a result of trial and error, we find the correlation between $a_{ji}$ and $L$ or the combination of $K_0$ and $L$, using the fitting formulas given by
\begin{gather}
  a_{ji} = a_{ji}^{(0)} + a_{ji}^{(1)} \left(\frac{y_{ji}}{100\ {\rm MeV}}\right) 
     + a_{ji}^{(2)} \left(\frac{y_{ji}}{100\ {\rm MeV}}\right)^2, \label{eq:aji} \\
  a_{j4} = a_{j4}^{(0)} + a_{j4}^{(1)} \left(\frac{y_{j4}}{100\ {\rm MeV}}\right)^{-1}
     + a_{j4}^{(2)} \left(\frac{y_{j4}}{100\ {\rm MeV}}\right)^{-2}, \label{eq:aj4} 
\end{gather}
for $j=0,1,2$ and $i=1,2,3$, where $y_{ji}$ denotes $L$ or the combination of $K_0$ and $L$ depending on $j$ and $i$, which are listed in Table~\ref{tab:a_yi}.
Here, we find such combinations by hand, assuming that the power is an integer.
In Fig.~\ref{fig:aji}, one can see how strong such correlations are, where the solid lines denote the fitting given by Eqs. (\ref{eq:aji}) and (\ref{eq:aj4}). The coefficients $a_{ji}^{(k)}$ for $j=0,1,2$ and $k=0,1,2$ in Eqs. (\ref{eq:aji}) and (\ref{eq:aj4}) are shown in Table~\ref{tab:a_ji^k}.
We emphasize that the coefficients, $a_{ji}$ and $a_{j4}$, can be expressed as a function of $y_{ji}$ or $y_{j4}$ independently of the uncertainties of $K_0$ and $L$, although we did not know the physical origin why such combinations shown in Table ~\ref{tab:a_ji^k} are suitable in the expression.
From Fig.~\ref{fig:aji}, one can observe that the correlation between $a_{ji}$ and $y_{ji}$ are weak for $i=2$ and 4. 

\begin{table}
\centering
\caption{$y_{ji}$ (combination of $K_0$ and $L$) for $i=1,2,3,4$ and $j=0,1,2$ in the fitting of $a_{0i}$, $a_{1i}$, and $a_{2i}$, given by Eqs. (\ref{eq:aji}) and (\ref{eq:aj4}). }
\begin{tabular}{cccc}
\hline\hline
       $i$           & $y_{0i}$ & $y_{1i}$ & $y_{2i}$    \\
\hline
   $1$  &  $(K_0^2L^3)^{1/5}$  &  $L$ & $L$  \\
   $2$  &  $(K_0^3L)^{1/4}$   & $(K_0^4L)^{1/5}$  & $(K_0^9L)^{1/10}$  \\
   $3$  &  $L$   & $L$  & $L$  \\
   $4$  &  $(K_0^3L)^{1/4}$   & $(K_0^2 L)^{1/3}$  & $(K_0^2 L)^{1/3}$  \\
\hline\hline
\end{tabular}
\label{tab:a_yi}
\end{table}

\begin{table*}
\centering
\caption{Coefficients $a_{ji}^{(k)}$ for $j=0,1,2$ and $k=0,1,2$ in the fitting of $a_{0i}$, $a_{1i}$, and $a_{2i}$ with Eqs.~(\ref{eq:aji}) and (\ref{eq:aj4}). }
\begin{tabular}{cccccccccc}
\hline\hline
         $i$         & $a_{0i}^{(0)}$ & $a_{0i}^{(1)}$  & $a_{0i}^{(2)}$ & 
                          $a_{1i}^{(0)}$ & $a_{1i}^{(1)}$ & $a_{1i}^{(2)}$  &
                          $a_{2i}^{(0)}$ & $a_{2i}^{(1)}$ & $a_{2i}^{(2)}$  \\
\hline
   $1$  & $0.0038337$ & $0.0073437$ & $-0.0068798$ & $0.083392$ & $-0.042113$ & $0.024719$ 
           & $-0.020817$ & $0.010515$ & $-0.0069116$   \\
   $2$  & $0.016178$ & $-0.019248$ & $0.0048505$ & $0.048374$ & $0.010393$ & $-0.0017163$ 
           & $-0.0088285$ & $-0.0040927$ & $0.00056104$   \\
   $3$  & $0.02413$ & $-0.053581$ & $0.033518$ & $0.1714$ & $-0.2740$ & $0.1446$ 
           & $-0.040238$ & $0.059085$ & $-0.028536$   \\
   $4$  & $0.0017045$ & $-0.0071152$ & $0.0062548$ & $0.018248$ & $-0.049949$ & $0.034971$
           & $-0.0045017$ & $0.012316$ & $-0.0086167$   \\
   \hline\hline
\end{tabular}
\label{tab:a_ji^k}
\end{table*}

\begin{figure*}[tbp]
\begin{center}
\includegraphics[scale=0.5]{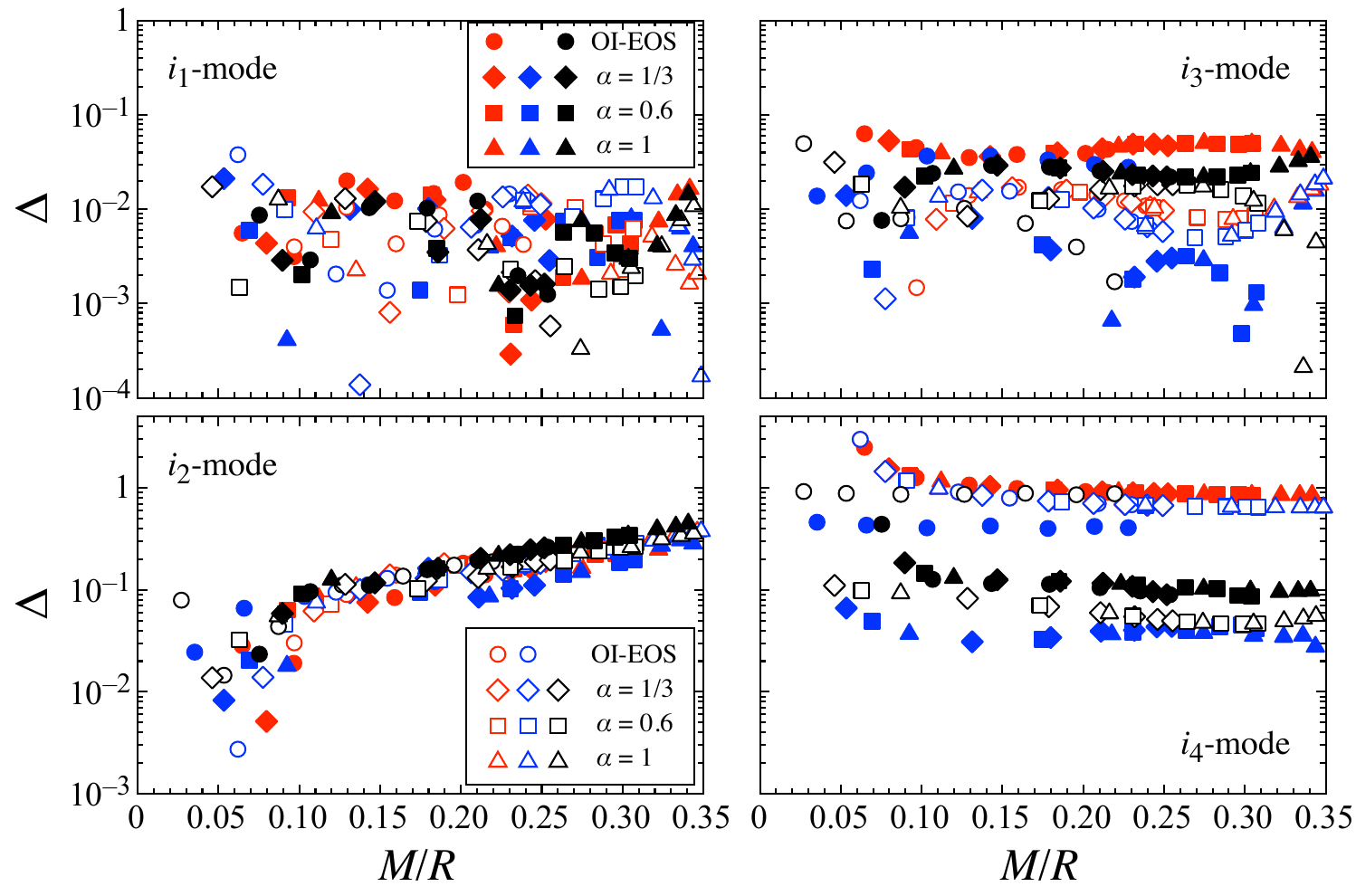}
\end{center}
\caption{
The relative deviation, $\Delta$, of the $i_i$-mode frequencies estimated with the empirical formulas from those determined via the eigenvalue problem, which is calculated with Eq. (\ref{eq:delta}). 
}
\label{fig:dfi}
\end{figure*}

Now, we can get a kind of empirical relations for expressing $f_iM$ as a function of $M/R$ and the nuclear saturation parameters, i.e., Eqs.~(\ref{eq:fit_i}), (\ref{eq:aji}), and (\ref{eq:aj4}). In Fig.~\ref{fig:dfi}, we show the relative deviation, $\Delta$, calculated by
\begin{equation}
  \Delta = |f-f_{\rm em}| / f,  \label{eq:delta}
\end{equation}
where $f$ and $f_{\rm em}$ denote the frequencies determined via the eigenvalue problem and those estimated with the empirical relations, respectively. The empirical relations we derived here tell us the $i_1$-mode frequencies within a few $\%$ accuracies for canonical neutron star models, but the $i_2$- and $i_4$-mode frequencies with only an order of magnitude due to the weak correlations with nuclear saturation parameters.

\subsection{Empirical relations for the $s$-modes}
\label{sec:ERs}

\begin{figure}[tbp]
\begin{center}
\includegraphics[scale=0.5]{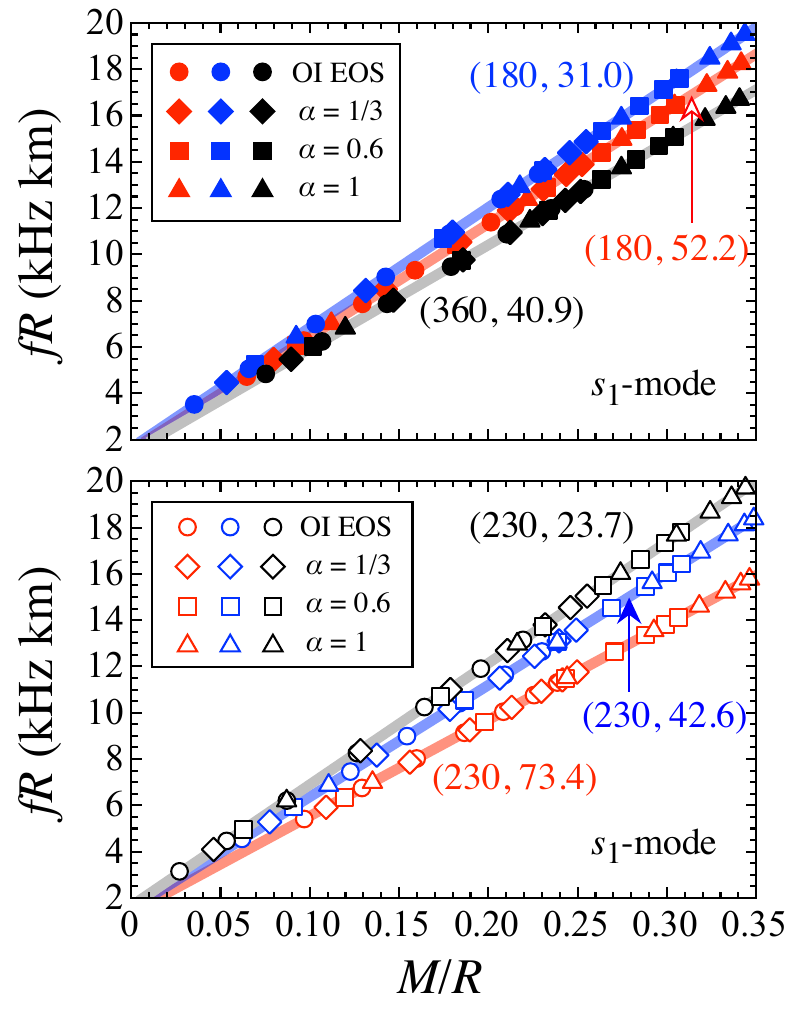}
\end{center}
\caption{
$f_{s_1}R$ for various stellar models are shown as a function of stellar compactness, where the circles, diamonds, squares, and triangles denote the results with stellar models constructed with original OI-EOS, OI-EOS connected to the EOS with $\alpha=1/3$, $0.6$, and $1$, respectively. The solid lines are the fitting given by Eq.~(\ref{eq:fit_s}) and we denote the values of $(K_0, L)$ on each line.
}
\label{fig:s1MR}
\end{figure}

\begin{figure*}[tbp]
\begin{center}
\includegraphics[scale=0.33]{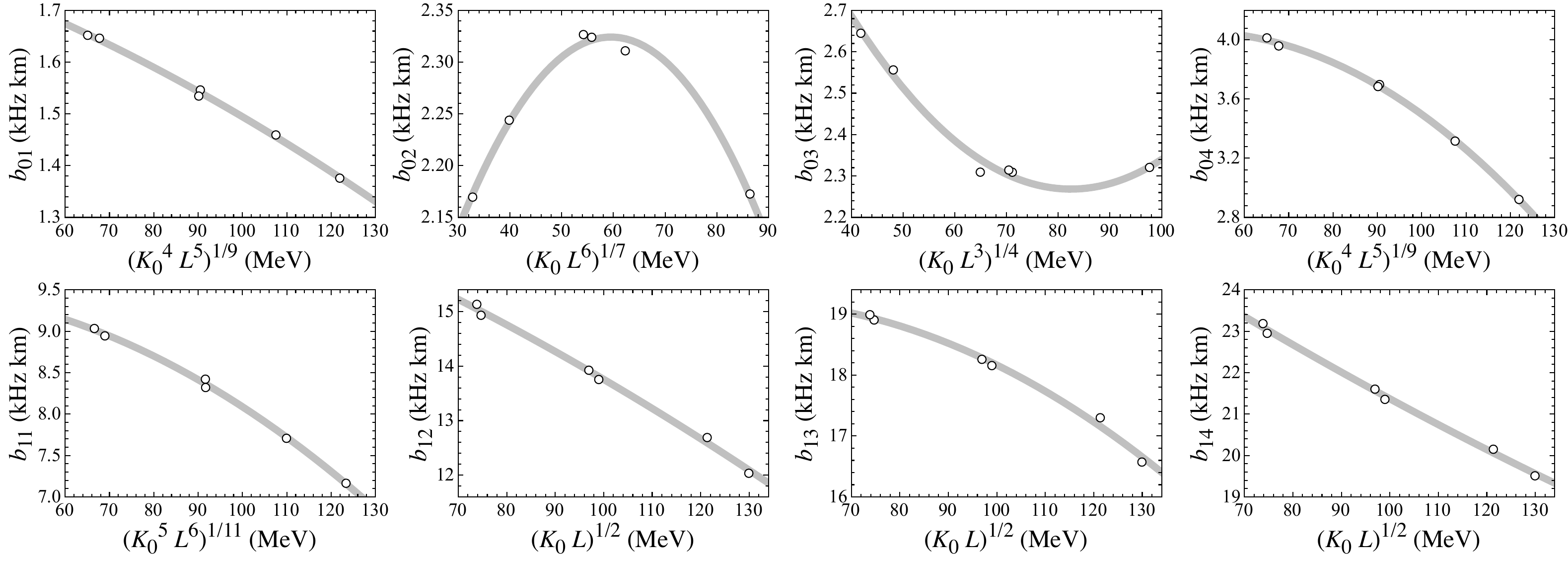}
\end{center}
\caption{
Coefficients, $b_{0i}$ and $b_{1i}$, in Eq.~(\ref{eq:fit_s}) are fitted as a function of $z_{0i}$ or $z_{1i}$ listed in Table~\ref{tab:b_zi}, which is the combination of $K_0$ and $L$ with using the fitting formulas given by Eqs. (\ref{eq:b0i}) and (\ref{eq:b1i}). The coefficients in Eqs.~(\ref{eq:b0i}) and (\ref{eq:b1i}), i.e., $b_{0i}^{(j)}$ and $b_{1i}^{(j)}$ for $j=0,1,2$, are listed in Table~\ref{tab:b_01i^j}.
}
\label{fig:b01i}
\end{figure*}

Next, we will see the behavior of the $s$-mode frequencies. In a similar way shown in the previous study \cite{Sotani23}, we confirm that $f_sR$ is well expressed as a linear function of $M/R$ as in Eq. (\ref{eq:fit_s}) without any exceptional instances, as shown in Fig.~\ref{fig:s1MR}, where the coefficients in Eq. (\ref{eq:fit_s}), i.e., $b_{0i}$ and $b_{1i}$, depend on the crustal properties. Again, through trial and error, we find that $b_{0i}$ and $b_{1i}$ can be expressed as a function of the combination of $K_0$ and $L$, using the fitting given by
\begin{gather}
  b_{0i} = b_{0i}^{(0)} + b_{0i}^{(1)} \left(\frac{z_{0i}}{100\ {\rm MeV}}\right) 
     + b_{0i}^{(2)} \left(\frac{z_{0i}}{100\ {\rm MeV}}\right)^2, \label{eq:b0i} \\
  b_{1i} = b_{1i}^{(0)} + b_{1i}^{(1)} \left(\frac{z_{1i}}{100\ {\rm MeV}}\right)
     + b_{1i}^{(2)} \left(\frac{z_{1i}}{100\ {\rm MeV}}\right)^2, \label{eq:b1i}
\end{gather}
where $z_{0i}$ and $z_{1i}$ are the specific combination of $K_0$ and $L$ depending on $i$, which are listed in Table~\ref{tab:b_zi}, and the coefficients in Eqs.~(\ref{eq:b0i}) and (\ref{eq:b1i}) are listed in  Table~\ref{tab:b_01i^j}. In Fig.~\ref{fig:b01i}, we show how the fitting of $b_{0i}$ and $b_{1i}$ works well with (\ref{eq:b0i}) and (\ref{eq:b1i}), where the marks denote the numerical values of $b_{0i}$ and $b_{1i}$ while the solid lines denote the fitting with Eqs.~(\ref{eq:b0i}) and (\ref{eq:b1i}).

\begin{table}
\centering
\caption{$z_{0i}$ and $z_{1i}$ (combination of $K_0$ and $L$) in the fitting of $b_{0i}$ and $b_{1i}$ expressed by Eqs. (\ref{eq:b0i}) and (\ref{eq:b1i}). }
\begin{tabular}{ccc}
\hline\hline
       $i$           & $z_{0i}$ & $z_{1i}$     \\
\hline
   $1$  &  $(K_0^4L^5)^{1/9}$  &  $(K_0^5L^6)^{1/11}$ \\
   $2$  &  $(K_0L^6)^{1/7}$   & $(K_0L)^{1/2}$  \\
   $3$  &  $(K_0L^3)^{1/4}$   & $(K_0L)^{1/2}$  \\
   $4$  &  $(K_0^4L^5)^{1/9}$   & $(K_0L)^{1/2}$  \\
\hline\hline
\end{tabular}
\label{tab:b_zi}
\end{table}

\begin{table}
\centering
\caption{Coefficients $b_{0i}^{(j)}$ and $b_{1i}^{(j)}$ for $j=0,1,2$ in the fitting of $b_{0i}$ and $b_{1i}$ expressed by Eqs. (\ref{eq:b0i}) and (\ref{eq:b1i}). }
\begin{tabular}{ccccccc}
\hline\hline
         $i$         & $b_{0i}^{(0)}$ & $b_{0i}^{(1)}$  & $b_{0i}^{(2)}$ & $b_{1i}^{(0)}$ & $b_{1i}^{(1)}$ & $b_{1i}^{(2)}$    \\
\hline
   $1$  & $1.8624$ & $-0.2316$ & $-0.1364$ & $9.4614$ & $0.7380$ & $-2.1114$   \\
   $2$  & $1.5724$ & $2.5259$ & $-2.1216$ & $17.8859$ & $-3.0316$ & $-1.0974$   \\
   $3$  & $3.8459$ & $-3.8269$ & $2.3209$ & $18.4672$ & $3.3390$ & $-3.6406$   \\
   $4$  & $3.5630$ & $2.0388$ & $-2.1046$ & $28.7034$ & $-8.3357$ & $1.0013$   \\
   \hline\hline
\end{tabular}
\label{tab:b_01i^j}
\end{table}

\begin{figure*}[tbp]
\begin{center}
\includegraphics[scale=0.5]{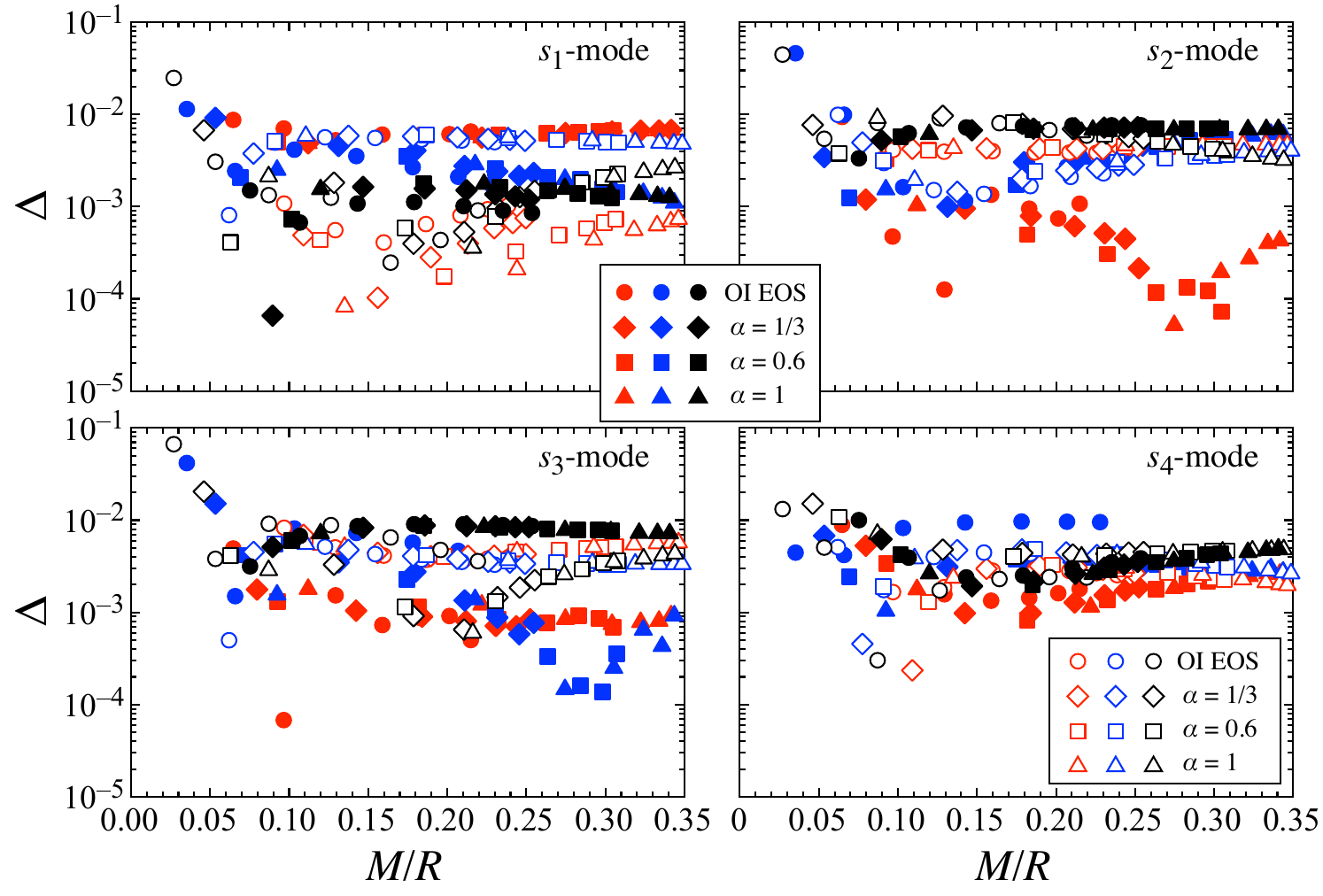}
\end{center}
\caption{
The relative deviation, $\Delta$, of the $s_i$-mode frequencies estimated with the empirical formulas from those determined via the eigenvalue problem, which is calculated with Eq. (\ref{eq:delta}). 
}
\label{fig:dfs}
\end{figure*}

Now, we get the empirical relations for $f_sR$ as a function of $M/R$ and the combination of the saturation parameters, given by Eqs.~(\ref{eq:fit_s}), (\ref{eq:b0i}), and (\ref{eq:b1i}). In Fig.~\ref{fig:dfs}, we show how our empirical relations work well, where the relative deviation, $\Delta$, is calculated with Eq. (\ref{eq:delta}). From this figure, one can observe that the empirical relations we derived can estimate the $s$-mode frequencies for a canonical neutron star within $\sim 1\%$ accuracy.

\section{Possibility for identification of higher-frequency QPOs}   
\label{sec:identification}

\begin{figure*}[tbp]
\begin{center}
\includegraphics[scale=0.6]{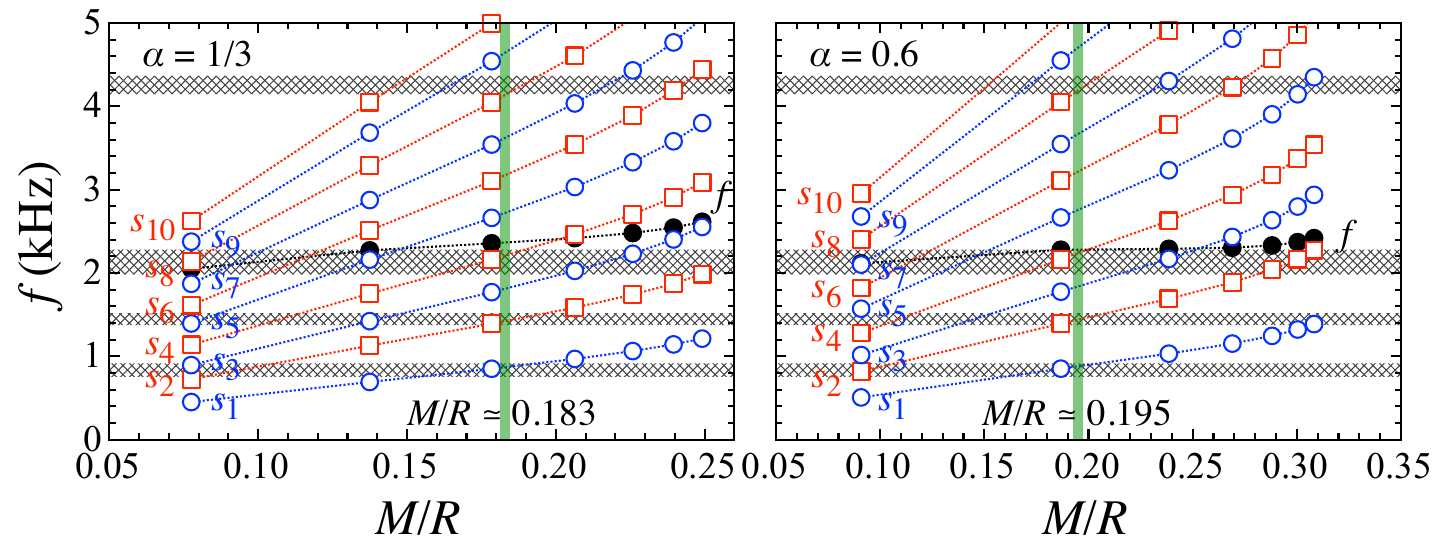}
\end{center}
\caption{
The $f$- and $s_i$-mode frequencies are shown as a function of the stellar compactness for the stellar models constructed with $(K_0, L)=(230, 42.6)$, together with the QPO frequencies observed in GRB 200415A \cite{CT21}. The left and right panels correspond to the results with $\alpha=1/3$ and 0.6, respectively. The horizontal shaded bands denote the observed QPO frequencies, i.e., $835.9^{-84.7}_{+77.3}$, $1443.7^{-68.7}_{+74.8}$, $2131.7^{-151.0}_{+148.2}$, and $4249.7^{-102.7}_{+116.0}$ Hz \cite{CT21}, while the marks denote the frequencies expected theoretically. The vertical lines denote the suitable stellar compactness for identifying the observed QPO frequencies with the $s_1$-, $s_2$-, $s_4$-, and $s_8$-modes, i.e., $M/R\simeq 0.183$ $(0.195)$ for $\alpha=1/3$ (0.6).
}
\label{fig:QPOs}
\end{figure*}

In the end, we consider the possibility of applying the $s$-mode for the QPO observations. The higher frequency QPOs, i.e., 836, 1444, 2132, and 4250 Hz, are found in GRB 200415A \cite{CT21}. Since GRB 200415A is classified as a magnetar giant flare, we have discussed the stellar model by identifying the observed QPOs with the overtones of crustal torsional oscillations~\cite{SKS23} with the same framework as for the identification of the QPOs observed in SGR 1806-20 and 1900+14. However, since the QPO frequencies observed in GRB 200415A are comparable to the other neutron star oscillations, such as the shear modes discussed in this paper, one may identify the observed QPO frequencies with other neutron star eigenmodes. In fact, we can find the possible identification with the shear oscillations. In Fig.~\ref{fig:QPOs}, we show the $f$- and $s_i$-mode frequencies for the stellar models with $(K_0, L)=(230, 42.6)$ are shown as a function of the stellar compactness, together with the QPO frequencies observed in GRB 200415A. The left and right panels correspond to the results with $\alpha = 1/3$ and 0.6, respectively. From this figure, one can identify the observed QPO frequencies with the $s_1$-, $s_2$-, $s_4$-, and $s_8$-mode frequencies, if $M/R\simeq 0.183$ for $\alpha = 1/3$ or $M/R\simeq 0.195$ for $\alpha =0.6$. One could discuss the crust properties via the identification of QPOs by the torsional oscillations~\cite{SKS23}, owing to the independence of their frequencies from the properties of the neutron star core, but it may not be so simple to discuss the crust properties via the identification by the shear oscillations as shown in Fig.~\ref{fig:QPOs}, because the shear oscillations depend not only the crust properties but also the core properties. Nevertheless, via the identification of the QPOs by the shear oscillations, one may derive a kind of constraint between the crust properties, core properties, and $M/R$. Such a possibility may be studied somewhere in the future.

\section{Conclusion}
\label{sec:Conclusion}

The existence of the crust elasticity can additionally excite the $i$- and $s$-modes. Because the crust thickness strongly depends on the stellar compactness and nuclear saturation parameters \cite{SIO2017}, one can expect that the $i$- and $s$-mode frequencies depend on such properties. In addition, since the $i$- and $s$-modes belong to the polar-type oscillations, they also depend on the properties of the neutron star core. Nevertheless, we have shown that $f_iM$ and $f_sR$ can be expressed as a function of the stellar compactness independently of the stiffness of core region \cite{Sotani23}. In this study, we further examine the dependence on the nuclear saturation parameters. As a result, we find a correlation between the coefficients in the fitting formulas with the stellar compactness and the nuclear saturation parameters, which gives us the empirical relations for expressing the $i$- and $s$-mode frequencies. Unfortunately, the empirical relations for the $i_2$- and $i_4$-modes do not work well, because the correlations with the nuclear saturation parameters are not so strong. Using the empirical relations we derived, one can estimate the $i_1$-modes within a few $\%$ and the $s$-mode within $\sim1\%$ accuracies for a canonical neutron star. We also show the possibility of identifying the higher QPO frequencies observed in GRB 200415A \cite{CT21} with the shear oscillations, as an alternative possibility instead of the torsional oscillations. Unlike the torsional oscillations, since the shear oscillations depend on not only the crust properties but also the stiffness of the core region, it may be more difficult to extract the physical information by identifying the QPO frequencies with the shear oscillations. Even so, via such an identification, one may be able to extract a kind of constraint on the relation between the crust properties, stiffness of the core region, and stellar compactness, which will be done somewhere in the future.

\acknowledgments
We are grateful to Kazuhiro Oyamatus for providing the EOS data and to Hajime Togashi for valuable comments.
This work is supported in part by Japan Society for the Promotion of Science (JSPS) KAKENHI Grant Numbers 
JP19KK0354  
and
JP21H01088,  
by FY2023 RIKEN Incentive Research Project,
and by Pioneering Program of RIKEN for Evolution of Matter in the Universe (r-EMU).

\appendix
\section{Dependence of $i$-mode frequencies on surface density }   
\label{sec:appendix_1}

\begin{figure}[tbp]
\begin{center}
\includegraphics[scale=0.5]{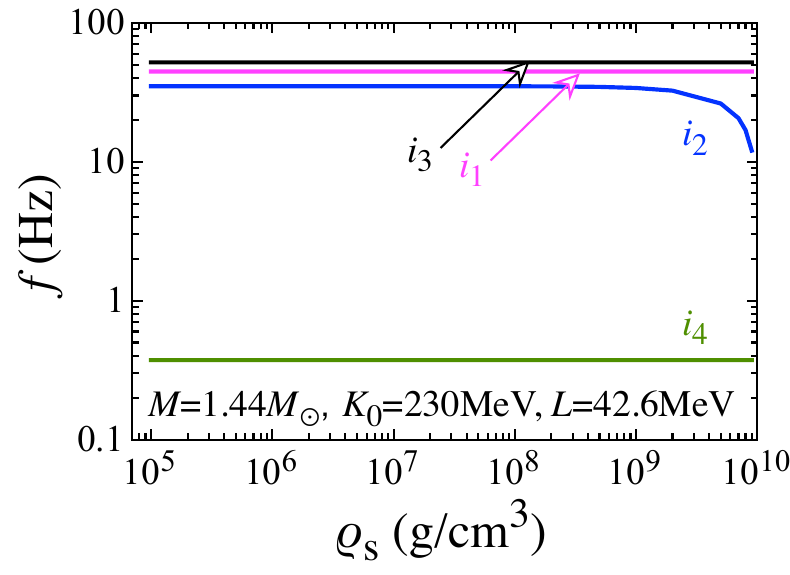}
\end{center}
\caption{
Dependence of the $i$-mode frequencies on the surface density, $\rho_s$, for the $1.44M_\odot$ neutron star model constructed with $K_0=230$ and $L=42.6$ MeV. The transition densities between the phases with and without elasticity are fixed as in the text. 
}
\label{fig:i-rhos}
\end{figure}

Unlike the $s$-mode oscillations, which are confined only inside the elastic region (see Fig.~6 in \cite{Sotani23}), the eigenfunctions of the $i$-modes exude outside the elastic region (see Fig.~\ref{fig:imode}). Thus, while the $i$-mode frequencies depend on the position of the interface (or the transition density at the interface), they may also depend on the surface density, although we fix it being $\rho_s=10^6$ g/cm$^3$ in this study. In this appendix, we check how the $i$-mode frequencies depend on the surface density. In Fig.~\ref{fig:i-rhos}, we show the $i$-mode frequencies as a function of $\rho_s$ for the $1.44M_\odot$ neutron star model constructed with $K_0=230$ and $L=42.6$ MeV. We note that the transition density between the envelope and crust is set to $10^{10}$ g/cm$^3$. From this figure, one can observe that the $i_1$-, $i_3$-, and $i_4$-modes are independent of the selection of surface density, while the $i_2$-mode significantly changes if the ratio of the surface density to the transition density between the envelope and crust becomes more than $10\%$.

\end{document}